\documentclass[pra,reprint,amsmath,amssymb,aps,a4paper,english,superscriptaddress]{revtex4-2}

\usepackage{graphicx}
\usepackage[utf8]{inputenc}
\usepackage{braket}
\usepackage{comment}
\usepackage{float}
\usepackage{siunitx}
\usepackage[colorlinks=true, linkcolor=blue, urlcolor=blue, citecolor=blue]{hyperref}
\usepackage{bbold}
\usepackage{mathtools}

\begin{document}

\title{Andreev bound states and supercurrent\\ in disordered spin--orbit-coupled nanowire SNS-junctions}

\author{Jonas Lidal}
\affiliation{Center for Quantum Spintronics, Department of Physics, Norwegian University of Science and Technology, NO-7491 Trondheim, Norway}
\affiliation{Norwegian National Security Authority, Norway}
\author{Jeroen Danon}
\affiliation{Center for Quantum Spintronics, Department of Physics, Norwegian University of Science and Technology, NO-7491 Trondheim, Norway}

\date{\today}

\begin{abstract}
We use a single-channel scattering formalism to derive general expressions for the Andreev bound-state energies and resulting current--phase relationship in a one-dimensional semiconductor-based SNS-junction, including arbitrarily oriented effective spin--orbit and Zeeman fields and taking into account disorder in the junction in by including a single scatterer with transmission probability $0 \leq T^2 \leq 1$, arbitrarily located in the normal region.
We first corroborate our results by comparing them to the known analytic limiting-case expressions.
Then we simplify our general result in several additional limits, including the case of the scatterer being at a specific location and the cases of small and large spin--orbit fields compared to the Zeeman splitting, assuming low transparency ($T\ll 1$).
We believe that our results could be helpful for disentangling the main spin-mixing processes in experiments on low-dimensional semiconductor-based SNS-junctions.
\end{abstract}

\maketitle

\section{\label{sec:intro}Introduction}
Hybrid superconductor--semiconductor devices have attracted much
attention in recent years, mostly due to the exotic spin physics that can emerge from the interplay of spin--orbit interaction (SOI), a strong Zeeman effect, and proximity-induced superconducting correlations~\cite{Bercioux2015,Cai2023}.
One potential application of such devices lies in the engineering of topological superconductivity~\cite{Sau2010,Oreg2010,Lutchyn2010,Lutchyn2018}, which is expected to come with localized Majorana-like states that could be useful for fault-tolerant topological quantum computation~\cite{Nayak2008,Sarma2015}.
The simultaneous breaking of inversion and time-reversal symmetry associated with the SOI and Zeeman splitting makes superconductor--semiconductor devices also natural candidates for exploring the so-called superconducting diode effect~\cite{Ando2020,Nadeem2023}, which is potentially useful in the context of low-dissipation electronics.
Furthermore, advancing this hybrid platform could result in a seamless integration of semiconductor- and superconductor-based technologies, for instance, enabling long-range qubit-qubit coupling of spin qubits via superconducting resonators~\cite{Burkard2020}.

One of the simplest of such hybrid devices one can make is the superconductor--normal--superconductor (SNS) junction, i.e., two superconducting regions that are connected via a normal (semiconducting) region.
A finite phase difference between the two superconductors typically results in a supercurrent through the junction, the so-called Josephson effect.
In the presence of SOI and a Zeeman splitting in the normal region, the Josephson effect can become asymmetric~\cite{Rasmussen2016}, yielding, e.g., a $\varphi_0$-junction (where the free energy is minimal at a finite phase difference and a finite supercurrent can run at zero phase difference) or supercurrent rectification (when the maximal supported supercurrent in the two directions of the junction is different).
This makes the semiconductor-based SNS-junction a versatile device that can be used for realizing superconducting diodes~\cite{Mazur2022,baumgartner2022,Baumgartner2022b,Davydova2022,Souto2022,Fominov2022,valentini2023,Ciaccia2023,Maiani2023}, to construct ``phase batteries''~\cite{Strambini2020,Mayer2020}, or to realize tunable $0$--$\pi$ junctions that could find use in superconducting qubit platforms~\cite{Kjaergaard2020}.
Moreover, they can potentially be tuned to a topological regime where the junction hosts Majorana bound states~\cite{Pientka2017,Fornieri2019}, making them a natural building block for two-dimensional arrays of Majorana states~\cite{Hell2017}.
Apart from all these envisioned applications, another potential use of SNS-junctions is as a probe to study the spin physics in the semiconductor, possibly revealing details about the interplay of the SOI and the Zeeman effect~\cite{Lidal2023} or about the occurrence of topological phase transitions~\cite{San-Jose2013,San-Jose2014,Murthy2020,Dominguez2022}.

Because of their interesting physics and the potential use of SNS-junctions, there is a considerable amount of theoretical literature investigating the (critical) current through such junctions~\cite{Golubov2004}.
In the absence of SOI and a Zeeman splitting, results have been derived for both short~\cite{Kulik1975,Beenakker1991,Beenakker1991b} and long~\cite{Kresin1986,Altshuler1987,Furusaki1991} junctions (compared to the superconducting coherence length).
When including spin mixing in the normal region, the problem becomes more complex~\cite{BuzdinReview,Buzdin1992,Buzdin2008,Bergeret2015,Bezuglyi2002,Barash2002,Dimitrova2006,Gogin2022}, and for higher-dimensional junctions it is not always possible to arrive at analytic results~\cite{Liu2010,Reynoso2008,Reynoso2012,Ness2022}.

A particularly simple approach that often yields explicit results is to consider a (quasi-)one-dimensional junction, where all transport takes place through a single or few channels.
In that case, the supercurrent can often be calculated analytically from the bound-state spectrum in the junction, which follows from solving a few-channel scattering problem.
Both for long~\cite{Krive2004} and short~\cite{Cheng2012} junctions, expressions have been derived that include the effect of SOI and a Zeeman splitting.
Taking also disorder into account is relatively straightforward in this setup: Both disorder at the SN-interfaces and inside the normal region can be modeled by including scaterrers with finite reflection probability, which resuled in analytic expressions for the supercurrent in certain limiting cases~\cite{Yokoyama2013,Yokoyama2014,Yokoyama2014b,Campagnano2015}.
These insights in the single-channel physics are not only crucial as building blocks for a deeper understanding of the multi-channel (two- or three-dimensional) case, but they are also experimentally relevant for hybrid devices based on one-dimensional semiconducting nanowires~\cite{Doh2005}.

In this paper, we develop the single-channel scattering approach of Refs.~\cite{Yokoyama2014,Yokoyama2014b} further, generalizing the analytic results to describe any desired combination of effective spin--orbit and Zeeman fields in the junction, including a single point-like scaterrer at an arbitrary location in the normal region to account for disorder.
We show how our expressions reduce to the known results in all limiting cases, and we demonstrate how additional limits can be investigated easily, yielding compact and insightful expressions for the supercurrent and critical current through the junction.
We believe that our results present a valuable contribution to the steadily growing understanding of hybrid superconductor--semiconductor devices and that they could be a useful asset for disentangling spin--orbit and Zeeman physics in experiments on low-dimensional SNS-junctions.

The rest of the paper is structured as follows.
In Sec.~\ref{sec:model} we introduce the setup we consider and the model we use to describe the system.
We then show how all ingredients of the model can be combined into a straightforward self-consistency equation for the bound-state energies in the long-junction limit.
In Sec.~\ref{sec:results} we present a general analytic solution of this equation, which can be used to find the (critical) current through the junction.
We first compare our general result to the known limiting-case expressions, and then both investigate new limits in the low-transparency case and present simple closed-form general expressions for fixed positions of the scatterer.
Finally, we present a short conclusion in Sec.~\ref{sec:conclusion}.
The full explicit expression of our analytic solution of the scattering problem is presented in an Appendix.

\section{\label{sec:model}Model}

\begin{figure}[t]
    \centering
    \includegraphics[width=0.45\textwidth]{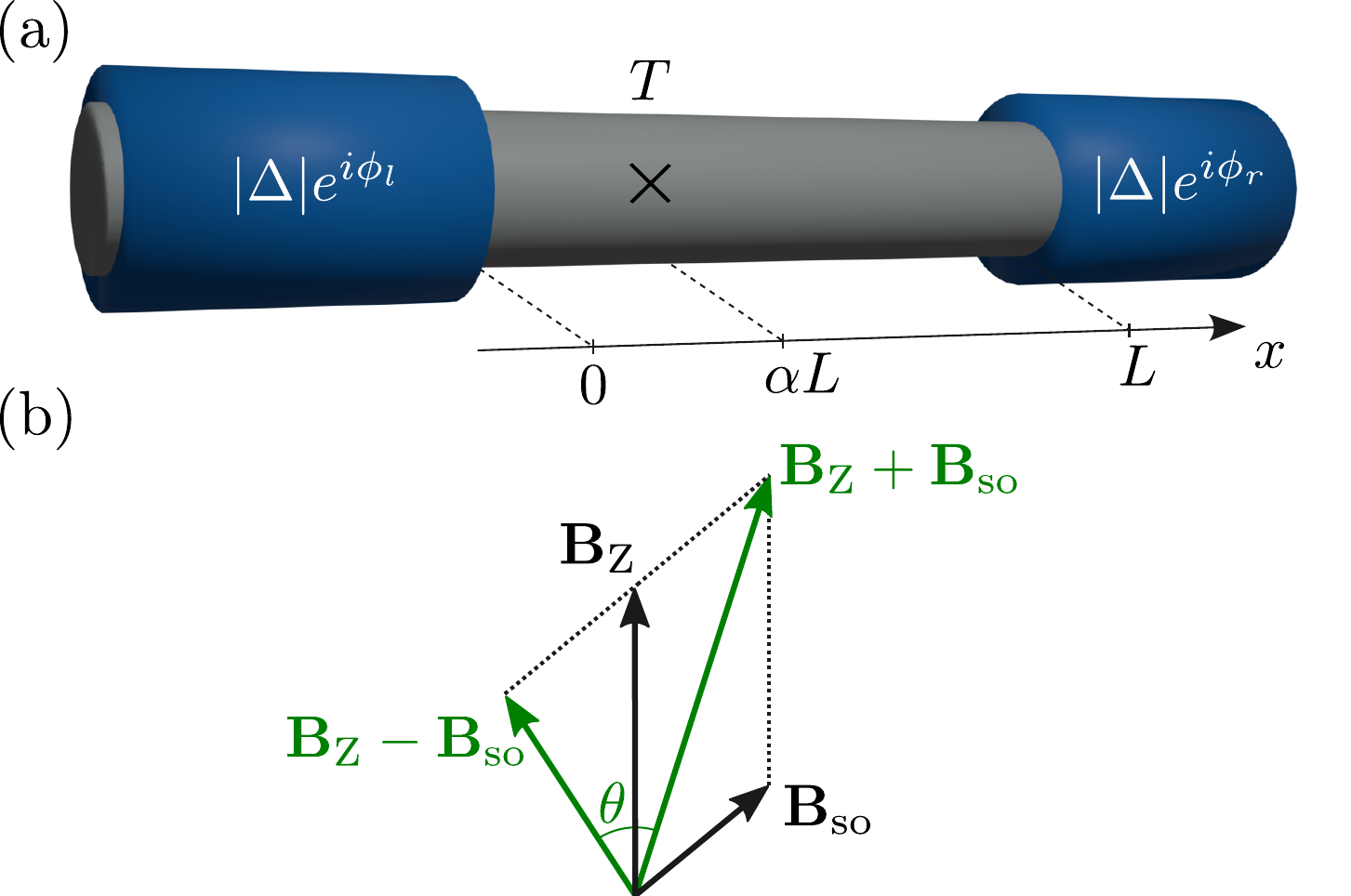}
    \caption{(a) Sketch of the setup considered.
    A one-dimensional semiconductor nanowire (gray) is connected to two superconductors (blue), separated by a distance $L$, which have a phase difference $\phi = \phi_l-\phi_r$.
    Disorder in the wire is modeled by a single scatterer with transmission probability $T$, located at position $x=\alpha L$.
    (b) The Zeeman and effective spin--orbit fields (black) are not necessarily parallel.
    The total fields the electrons traveling in the two directions experience are their sum and difference fields (green).}
    \label{fig:device}
\end{figure}

\textit{Setup}---We consider a one-dimensional single-channel semiconductor nanowire connected to two superconducting contacts that are separated by a distance $L$, as sketched in Fig.~\ref{fig:device}(a).
Due to the proximity effect, the two ends of the wire will thus acquire a finite superconducting pairing potential, which we assume to be of equal strength on the two sides.
The wire can have significant spin--orbit coupling and we also include an external magnetic field that gives rise to a Zeeman splitting inside the wire.
We model the disorder in the junction in the simplest way possible, i.e., by including a single scatterer, which is characterized by a transmission amplitude $T$ and is located at the position $x = \alpha L$ with $0\leq \alpha \leq 1$.

We describe the electrons in the wire using a Bogoliubov--de Gennes Hamiltonian, $H=\frac{1}{2} \int dx\, \Psi^\dagger {\cal H} \Psi$, where $\Psi = [\psi_{\uparrow}(x),\psi_{\downarrow}(x),\psi_{\downarrow}^\dagger(x),-\psi_{\uparrow}^\dagger(x)]^T$ with $\psi_{\uparrow(\downarrow)}(x)$ being the annihilation operator for an electron with spin up(down) at position $x$.
We further use
\begin{equation}
 {\cal H} = \left(\begin{array}{c c}{{\hat {\cal H}_{0}}}&{{\hat{\Delta}}}\\ {{{\hat{\Delta}}^{\dagger}}}&{{-\mathcal{T} \hat {\cal H}_{0}\mathcal{T}^{\dagger}}}\end{array}\right),
\end{equation}
where $\hat {\cal H}_0$ and $\hat{\Delta}$ are $2 \times 2$ matrices in spin space and $\mathcal{T}$ is the time-reversal operator.
The Hamiltonian $\hat{\cal H}_0$ reads formally as
\begin{equation}\label{eq:ham}
    \hat {\cal H}_0 = -\frac{\hbar^2\partial_x^2}{2m} - \mu + V_0 \delta\left(x-\alpha L\right) +
    ({\bf{B}}_{\text{Z}} - i\partial_x \boldsymbol\alpha_{\rm so}) \cdot {\boldsymbol{\sigma}},
\end{equation}
where $V_0$ determines the strength of the scattering potential, ${\bf B}_{\rm Z}$ is the Zeeman field, and the vector $\boldsymbol \alpha_{\rm so}$ characterizes the spin--orbit coupling.
The vector $\boldsymbol \sigma = \{\sigma_x,\sigma_y,\sigma_z\}$ consists of the three Pauli matrices acting in spin space.
In our Nambu basis, the off-diagonal block in ${\cal H}$ becomes $\hat\Delta = \Delta(x)\mathbb{1}$, where we assume the pairing potential to be
\begin{equation}
    \Delta(x) = \begin{cases}
        \Delta_0 e^{i\phi_l}, \quad & \text{for } x<0,\\
        \Delta_0 e^{i\phi_r}, \quad & \text{for } x>L,\\
        0, \quad & \text{for } 0 \leq x \leq L,
    \end{cases}
\end{equation}
which describes the induced superconductivity in the two ends of the wire, $\phi_{l(r)}$ being the phase of the left(right) superconducting contact.

The goal is to find the supercurrent through the junction $I(\phi)$ as a function of the phase difference $\phi = \phi_l - \phi_r$ between the two superconductors.
Assuming the short-junction limit, i.e. the length of the junction being much smaller than the coherence length in the normal part, $L\ll\xi_N \sim \hbar v_{\rm F}/\Delta_0$, with $v_{\rm F}$ the Fermi velocity, and for simplicity focusing on the limit of zero temperature, the supercurrent follows as~\cite{Beenakker1991}
\begin{equation}\label{eq:supercurrent}
    I(\phi)={\frac{2e}{\hbar}}{\sum_{n}}^{\prime}{\frac{d E_{n}}{d\phi}},
\end{equation}
where the $E_n$ are the energies of the subgap Andreev bound states in the junction and the prime in the sum indicates that we include negative Andreev levels only.
This expression gives the supercurrent in the ground-state configuration of the junction, thus assuming that it had enough time to completely thermalize.
When the phase difference $\phi$ varies over time, then the supercurrent can be different depending on whether the variation is slow enough to keep the system (approximately) in its ground state continuously or not~\cite{Yokoyama2014}; this also leads to different maximal (critical) supercurrents that can be supported in thermalized and non-thermalized junctions.
Here, we will assume for simplicity that the system is always in its ground state, in which case Eq.~(\ref{eq:supercurrent}) gives the correct current--phase relationship.

\begin{figure}[t]
    \centering
    \includegraphics[width=0.45\textwidth]{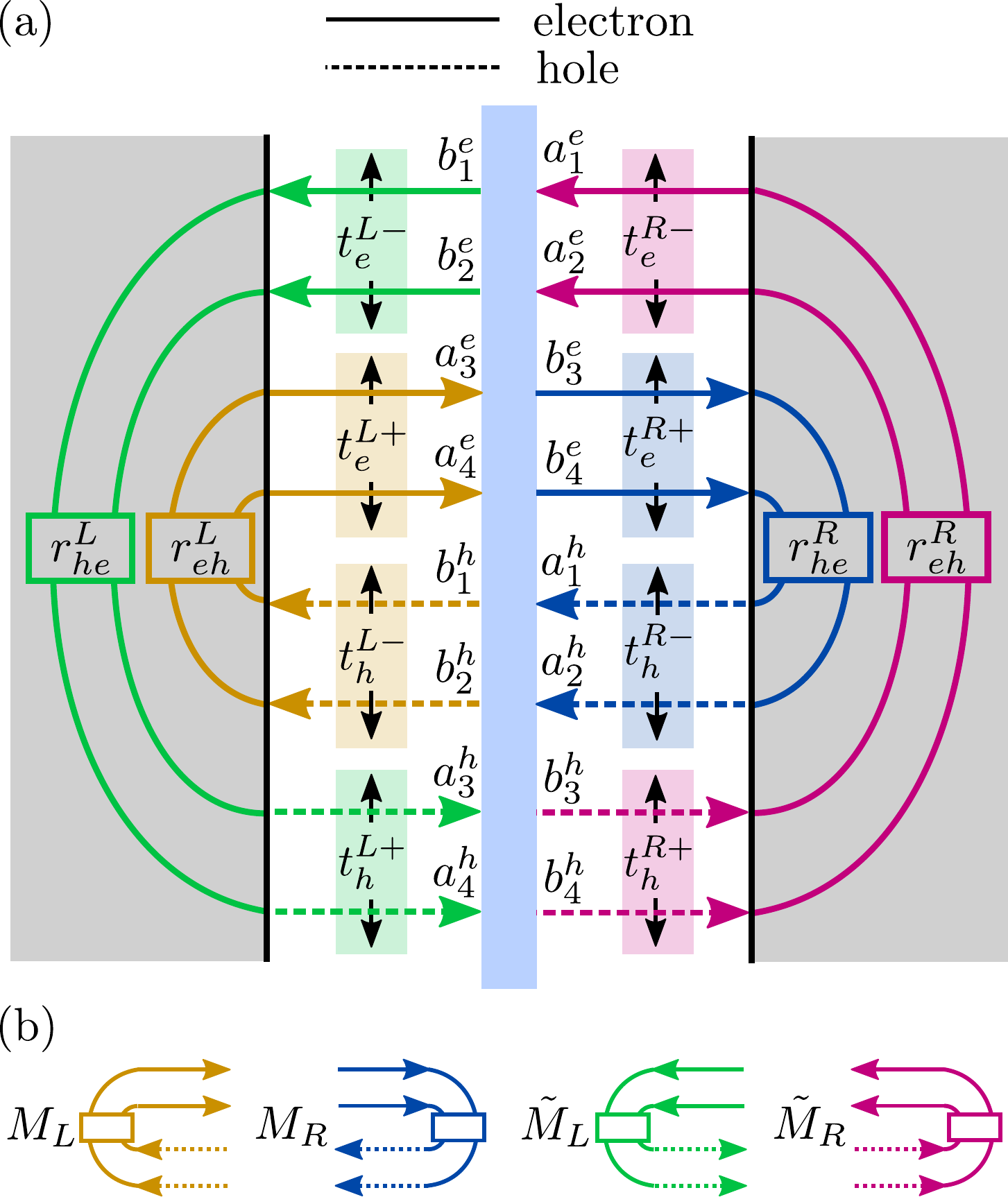}
    \caption{(a) Diagram summarizing all scattering processes.
    The vertical blue shaded region represents the scatterer, where electrons and holes can be transmitted or reflected.
    The horizontal arrows depict the free-propagation amplitudes of the electrons (solid) and holes (dashed), see Eq.~(\ref{eq:transfer}), where the small vertical arrows indicate the two spin directions in the local eigenbasis.
    Andreev reflection at the interfaces is represented by the bent amplitudes in the gray shaded areas, connecting electron to hole amplitudes and including spin mixing due to our choice of basis in the normal region.
    (b) Combinations of scattering processes that are combined into the matrices $M_{L,R}$ and $\tilde M_{L,R}$, see Eqs.~(\ref{eq:ml},\ref{eq:mr}).
    }
    \label{fig:amplitudes}
\end{figure}

We calculate the subgap Andreev levels, $E_n$, using a quantization argument, where we describe the propagation of subgap electrons and holes in the system in terms of a scattering matrix.
We split our scattering description into the following parts, as illustrated in Fig.~\ref{fig:amplitudes}(a):
(i) free propagation of electrons and holes on both sides of the scatterer, pictured with the straight solid and dashed arrows, respectively (the small black vertical arrows label the two different spin species), (ii) Andreev reflection at the SN-interfaces on the left and right ends of the wire, illustrated by the curves in the grey regions connecting incoming electrons to outgoing holes and vice versa, and finally (iii) elastic scattering by the impurity (light blue strip) at position $\alpha L$.
Below we will consider these three ingredients in detail and explain how we combine them to find the energies of the Andreev bound states.

\textit{Free propagation}---We start by assuming translational invariance in the two non-proximitized parts of the wire on both sides of the scatterer, i.e., for $0\leq x < \alpha L$ and $\alpha L < x \leq L$, which allows us to write for these regions
\begin{align}
    \hat{\cal H}_0^{L,R} = \frac{\hbar^2k^2}{2m} - \mu +
    ({\bf{B}}_{\text{Z}} +k \,\boldsymbol\alpha_{\rm so}) \cdot {\boldsymbol{\sigma}},
\end{align}
in terms of the wave vector $k$.
Assuming that $\Delta_0 \ll \mu$ and anticipating that we will focus on the subgap physics of the junction, we linearize this Hamiltonian around the Fermi wave vector,
\begin{align}
    \hat{\cal H}_{0,\eta}^{L,R} \approx E_{\rm F} + \hbar v_{\rm F} (\eta k-k_{\rm F}) +
    ({\bf{B}}_{\text{Z}} +\eta {\bf B}_{\rm so}) \cdot {\boldsymbol{\sigma}},
\end{align}
where $\eta =\pm 1$ corresponds to electrons moving in the $\pm \hat x$-direction and $v_{\rm F}$ is the Fermi velocity.
We also introduced the effective spin--orbit field ${\bf B}_{\rm so} = k_{\rm F}\boldsymbol\alpha_{\rm so}$, dropping the small correction $(k-k_{\rm F})\boldsymbol\alpha_{\rm so}$.

We will describe the free propagation always in a ``local'' eigenbasis of spin, i.e., always setting the spin quantization axis along ${\bf{B}}_{\text{Z}} + {\bf B}_{\rm so}$ and ${\bf{B}}_{\text{Z}} - {\bf B}_{\rm so}$ depending on the direction of propagation of the particle under consideration [see Fig.~\ref{fig:device}(b)].
Using all the conventions mentioned above, we then find that the transfer matrices that describe the propagation (see Fig.~\ref{fig:amplitudes}) can be written as
\begin{equation}
    t_{e,h}^{L\pm} = e^{i\sigma_z \chi_\pm^{e,h} \alpha}
    \quad\text{and}\quad
    t_{e,h}^{R\pm} = e^{i\sigma_z \chi_\pm^{e,h} (1-\alpha)},
    \label{eq:transfer}
\end{equation}
using the phases $\chi^e_\pm = -V_\pm L/\hbar v_{\rm F} + EL/\hbar v_{\rm F} + k_{\rm F}L$ and $\chi^h_\pm = -V_\pm L/\hbar v_{\rm F} - EL/\hbar v_{\rm F} - k_{\rm F}L$ with $V_\pm = |{\bf B}_{\rm Z} \pm {\bf B}_{\rm so}|$, $E$ the energy of the bound state, and $k_{\rm F}$ the Fermi wave number.
Since we work in the short-junction limit, we anticipate that the contributions $EL/\hbar v_{\rm F}$ can be neglected.
Furthermore, we see from Fig.~\ref{fig:amplitudes} that for the case of perfect Andreev reflection at the SN-interfaces the phase factors $\pm k_{\rm F} L$ cancel since they always appear with opposite signs in Andreev-reflected pairs of amplitudes.
Under our assumptions, it thus suffices to focus on the phases $V_\pm$, and we can drop the indices $e,h$ in Eq.~(\ref{eq:transfer}).

\textit{Andreev reflection}---The price to pay for aligning the spin quantization axis with the local fields is that all reflection processes become effectively spin-mixing.
The Andreev reflection coefficients thus become matrices that include a spin rotation that depends on the misalignment of the fields ${\bf B}_{\rm Z} \pm {\bf B}_{\rm so}$.
We find that the four matrices can be written as
\begin{align}
    r_{he}^L = {} & {} e^{-i\alpha} R(-\theta) e^{-i\phi_l}, &
    r_{eh}^L = {} & {} e^{-i\alpha} R(-\theta) e^{i\phi_l}, \\
    r_{he}^R = {} & {} e^{-i\alpha} R(\theta) e^{-i\phi_r}, &
    r_{eh}^R = {} & {} e^{-i\alpha} R(\theta) e^{i\phi_r},
\end{align}
where $\alpha = \arccos(E/\Delta_0)$ and
\begin{equation}
    R(\theta) =
    \begin{pmatrix}
    \cos (\frac{\theta}{2}) & \sin (\frac{\theta}{2}) \\
    -\sin (\frac{\theta}{2}) & \cos (\frac{\theta}{2})
    \end{pmatrix},
\end{equation}
implements a spin rotation over the angle
\begin{equation}
    \theta = \arccos \left( \frac{B_{\rm Z}^2 - B_{\rm so}^2}{V_+V_-} \right),
\end{equation}
i.e., the angle between ${\bf B}_{\rm Z} + {\bf B}_{\rm so}$ and ${\bf B}_{\rm Z} - {\bf B}_{\rm so}$, as indicated in Fig.~\ref{fig:device}(b).

We can now construct the matrices
\begin{align}
    M_L = {} & {} t_{e}^{L+} r_{eh}^L t_{h}^{L-}, &
    \tilde{M}_L = {} & {} t_{h}^{L+} r_{he}^L t_{e}^{L-},\label{eq:ml}\\
    M_R = {} & {} t_{h}^{R-} r_{he}^R t_{e}^{R+}, &
    \tilde{M}_R = {} & {} t_{e}^{R-} r_{eh}^R t_{h}^{R+},\label{eq:mr}
\end{align}
which fully describe all propagation on the left and right side of the scatterer, see also Fig.~\ref{fig:amplitudes}(b).

\textit{Impurity scattering}---Finally we consider the scatterer, which we assume to be spin-independent and particle--hole symmetric.
Defining the vector of incoming and outgoing electron amplitudes as ${\bf a}_e = (a^e_1,a^e_2,a^e_3,a^e_4)^T$ and ${\bf b}_e = (b^e_1,b^e_2,b^e_3,b^e_4)^T$, respectively, and similarly for the incoming and outgoing hole amplitudes [see Fig.~\ref{fig:amplitudes}(a)], we thus assume that the effect of the scatterer is to relate the amplitudes as ${\bf b}_{e,h} = S {\bf a}_{e,h}$ with
\begin{equation}
    S =
    \begin{pmatrix}
        T\,\mathbb{1}  & -\sqrt{1-T^2}\, R(\theta) \\
        \sqrt{1-T^2}\, R(-\theta) & T\,\mathbb{1}
    \end{pmatrix},
\end{equation}
where $T^2$ is the transmission probability of the scatterer.
The rotation matrices included with the probability amplitudes for reflection again take care of the misalignment of the spin bases for left- and right-moving particles.

\textit{Bound state}---We now see that, with the notation introduced before, we can also relate
\begin{equation}
    {\bf a}_h =
    \begin{pmatrix}
        0 & M_R \\ \tilde M_L & 0
    \end{pmatrix}
    {\bf b}_e,
    \quad
    {\bf a}_e = 
    \begin{pmatrix}
        0 & \tilde M_R \\ M_L & 0
    \end{pmatrix}
    {\bf b}_h.
    \label{eq:eigen}
\end{equation}
Combining Eq.~(\ref{eq:eigen}) with the relation ${\bf b}_{e,h} = S {\bf a}_{e,h}$ we find an eigenvalue problem that defines the bound states in the junction.
It thus follows that the equation
\begin{equation}
    \text{Det} \left[ 1 -  
    S
    \begin{pmatrix}
        0 & \tilde M_R \\ M_L & 0
    \end{pmatrix}
    S
    \begin{pmatrix}
        0 & M_R \\ \tilde M_L & 0
    \end{pmatrix}
    \right] = 0,\label{eq:det}
\end{equation}
in general constitutes a self-consistency equation for the energy $E$ of the bound state.
In our case, however, where $E$ only appears in the prefactor of all Andreev reflection coefficients, it is possible to write down a closed-form expression for the allowed energies.

\section{Results}\label{sec:results}

\begin{figure*}[t!]
    \centering
    \includegraphics[width=\textwidth]{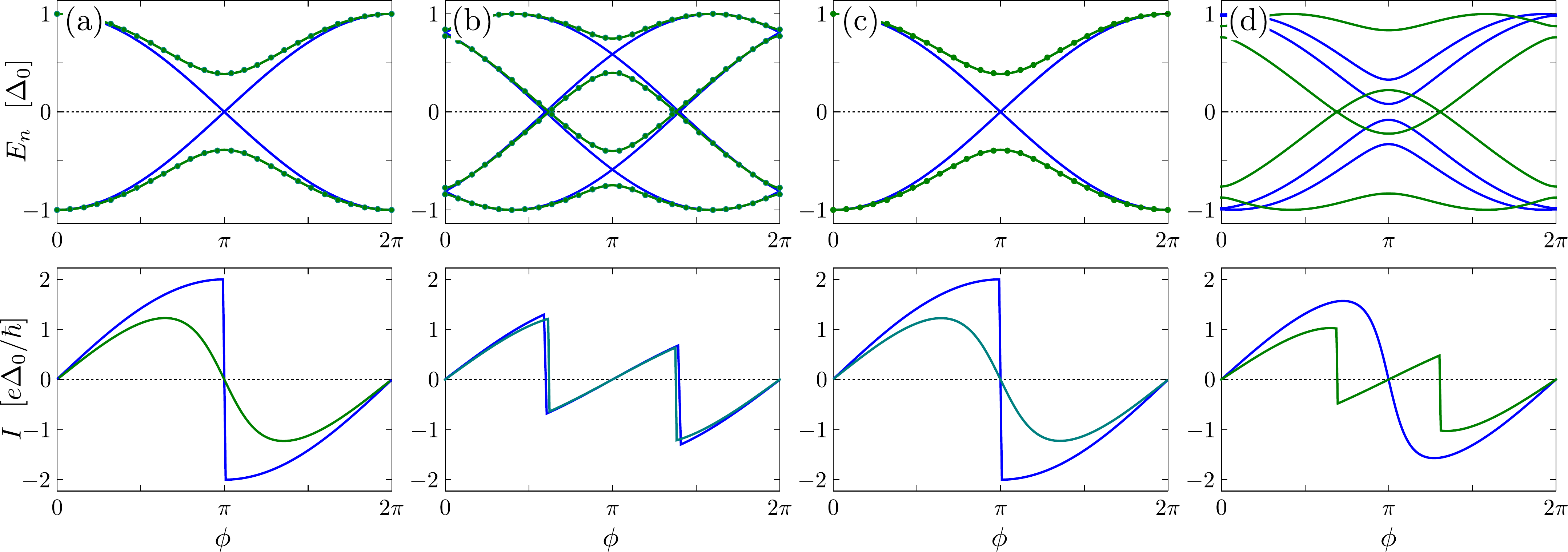}
    \caption{Calculated bound-state energies (top) and supercurrent (bottom) using Eqs.~(\ref{eq:energy}) and (\ref{eq:current}), respectively.
    The green dots present known results from the literature, for comparison.
    We set $\alpha = 1/\sqrt 2$ throughout and further we used:
    (a) $\chi_{\rm Z}=\chi_{\rm so}=0$ and $T=1$ (blue) or $T=\sqrt{0.85}$ (green);
    (b) $\chi_{\rm Z} = 0.2\,\pi$, $\chi_{\rm so} = 0$, and $T=1$ (blue) or $T=\sqrt{0.95}$ (green);
    (c) $\chi_{\rm Z} = 0$, $\chi_{\rm so} = 0.2\,\pi$, and $T=1$ (blue) or $T=\sqrt{0.85}$ (green);
    (d) $\chi_{\rm Z} = 0.2\,\pi$, $\chi_{\rm so} = 0.8\,\pi$, $T=\sqrt{0.95}$, and ${\bf B}_{\rm Z} \perp {\bf B}_{\rm so}$ (blue), and $\chi_{\rm Z} = \chi_{\rm so} = 0.2\,\pi$, $T=\sqrt{0.85}$, and an angle of $\pi/4$ between the two fields.}
    \label{fig:general}
\end{figure*}

Using the explicit expressions for the matrices given above, it is possible to rewrite Eq.~(\ref{eq:det}) in the form
\begin{equation}\label{eq:energy}
        E(\phi) = \pm \frac{\Delta_0}{\sqrt 2} \sqrt{v\pm \sqrt{w}},
\end{equation}
where
\begin{align}
    v = {} & {} 1 - q_1(T) + q_2(T) \cos (\phi), \\
    w = {} & {} \tfrac{1}{2} + [q_1(T) - q_2(T) \cos(\phi)]^2\nonumber\\
    {} & {} \hspace{1em} - q_3(T) - q_4(T) \cos(\phi) - T^4 \cos^2(\phi),
\end{align}
revealing the functional dependence of the energies on the phase difference $\phi$.
The coefficients $q_i$ are cumbersome, but straightforward functions of $T$ as well as $\theta$, $\alpha$, and $\chi_\pm$, and are given explicitly in App.~\ref{app:exp}.
Although the lengthy expressions for the $q_i$ make the general result for $E(\phi)$ slightly obscure at first sight, we believe that it is in fact very useful in several regards:
Firstly, the expression (\ref{eq:energy}) is written in such a way that the positive- and negative-energy solutions can be distinguished easily, allowing for an explicit analytic evaluation of the zero-temperature equilibrium supercurrent via (\ref{eq:supercurrent}),
\begin{equation}
    I(\phi) = -\frac{e\Delta_0}{\hbar} \frac{1}{\sqrt 2} \sum_{\eta = \pm 1}
    \frac{(\partial_\phi v) + \frac{1}{2}\eta (\partial_\phi w)/\sqrt w}{\sqrt{v +\eta \sqrt w}},\label{eq:current}
\end{equation}
where the differentiation with respect to $\phi$ yields
\begin{align}
    \partial_\phi v = {} & {} -q_2(T) \sin (\phi), \\
    \partial_\phi w = {} & {} [2 q_1(T)q_2(T)+q_4(T)]\sin(\phi)  \nonumber\\
    {} & {} \hspace{1em}  + [T^4 - q_2(T)^2] \sin (2\phi).
\end{align}
Secondly, using the explicit expressions given in the Appendix, it is straightforward to investigate limiting cases, e.g., of small or large $|{\bf B}_{\rm Z}|/|{\bf B}_{\rm so}|$ or $T$, or to simplify the general result for a specific location for the scatterer, such as $\alpha = 0$ or $\alpha = 1/2$.

Below we will present our results, based on Eq.~(\ref{eq:energy}), in the following way.
First, we will explore the general expressions for the bound-state energies and supercurrent, investigate their dependence on our model parameters, and compare the results with the expected behavior.
Then we will consider the limits of $T\ll 1$ and $T=1$, where it is straightforward to write down analytic expressions for the critical current $I_c = {\rm max}_\phi I(\phi)$, which can be connected to the magnitude and relative orientation of the fields ${\bf B}_{\rm Z}$ and ${\bf B}_{\rm so}$.
Finally we will illustrate how fixing $\alpha$ to a specific value---we will pick $\alpha = 0$ and $\alpha=1/2$---can allow to greatly simplify the general result, yielding digestible expressions for the bound-state energies and the supercurrent through the junction.

\subsection{General case}

In Fig.~\ref{fig:general} we plot the four bound-state energies (top row) and supercurrent through the junction (bottom row) as a function of the phase difference $\phi$, as given by Eqs.~(\ref{eq:energy}) and (\ref{eq:current}), respectively.
In Fig.~\ref{fig:general}(a) we set both phases $\chi_{\rm Z} \equiv |{\bf B}_{\rm Z}|L/\hbar v_{\rm F}$ and $\chi_{\rm so} \equiv |{\bf B}_{\rm so}|L/\hbar v_{\rm F}$ to zero, to see if our expressions reproduce the usual results.
The blue curves are for $T = 1$ and the green curves for $T = \sqrt{0.85}$, and in both cases we set $\alpha = 1/\sqrt 2$ (which we keep at this value throughout the whole Figure).
We see that the energies indeed follow the expected $E(\phi) = \pm \Delta_0 \cos(\phi/2)$ for the fully transparent junction and open up an anticrossing around $\phi = \pi$ for lower transparency.
The green dots show the energies given by Beenakker's well-known expression $E(\phi) = \pm\Delta_0 \sqrt{1-T^2 \sin(\phi/2)^2}$ for a short single-channel Josephson junction~\cite{Beenakker1991}.
In Fig.~\ref{fig:general}(b) we add a finite Zeeman field, setting $\chi_{\rm Z} = 0.2\,\pi$, while keeping $\chi_{\rm so} = 0$.
As expected, the doubly (spin) degenerate levels of Fig.~\ref{fig:general}(a) now split.
For $T=1$ (blue curves), the effect of the Zeeman field is simply an extra phase shift that is acquired by electrons and holes crossing the junction, opposite for the two spin directions~\cite{Cheng2012}.
The bound-state energies thus follow $E(\phi) = \pm \Delta_0 \cos( \phi/2 \pm \chi_{\rm Z})$, and the system can undergo a $0$--$\pi$ transition as the Zeeman field is increased~\cite{BuzdinReview,FuldeFerrell1964,LarkinOvchinnikov,Buzdin1982,Buzdin1992,Ryazanov2001,Cai2023}.
In the presence of a finite scattering probability, level anticrossings will emerge.
This is illustrated by the green curve, where we set $T = \sqrt{0.95}$.
The green dots show the energies as given by the simple expression derived in Ref.~\cite{Yokoyama2013} [Eqs.~(23,24), valid in the limit $\chi_{\rm so} =0$], which agree with our more general expression.
In Fig.~\ref{fig:general}(c) we show the opposite case of finite spin--orbit coupling, $\chi_{\rm so} = 0.2\,\pi$ and zero Zeeman splitting, where the transparency is set to $T=1$ (blue) and $T=\sqrt{0.85}$ (green).
These results are identical to the effectively spinless case shown in Fig.~\ref{fig:general}(a), where $\chi_{\rm Z} = \chi_{\rm so} = 0$.
Indeed, in the short-junction limit, phase shifts purely due to spin--orbit coupling cancel, and deviations from the spinless case appear only in higher order in $\Delta_0 \tau_{\rm dwell}/\hbar$, where $\tau_{\rm dwell}$ is the dwell time of the current-carrying particles in the junction~\cite{Chtchelkatchev2003,Beri2008}.
Finally, in Fig.~\ref{fig:general}(d) we explore more general parameter settings:
In blue we show the case $\chi_{\rm Z} = 0.2\,\pi$, $\chi_{\rm so} = 0.8\,\pi$, and $T=\sqrt{0.95}$, where the fields ${\bf B}_{\rm Z}$ and ${\bf B}_{\rm so}$ were assumed perpendicular, and in green the case $\chi_{\rm Z} = \chi_{\rm so} = 0.2\,\pi$, and $T=\sqrt{0.85}$, now using an angle of $\pi/4$ between the two fields.
In all columns, the lower panel shows the corresponding current--phase relationship at zero temperature, which in principle follows from Eq.~(\ref{eq:supercurrent}) as the phase derivative of the sum of all negative bound-state energies.
We emphasize again that, due to the form of Eq.~(\ref{eq:energy}), our expression for the supercurrent (\ref{eq:current}) is written in a closed form and does not require a separate determination of which of the four bound-state energies are negative.
We see that we never find an anomalous Josephson current (a finite supercurrent at $\phi=0$), which is already clear from the observation that the bound-state spectrum is symmetric in $\phi$, i.e., $E(\phi) = E(-\phi)$~\cite{Yokoyama2013}.

The results presented in this Section show how Eqs.~(\ref{eq:energy}) and (\ref{eq:current}) can be straightforwardly used to calculate the bound-state spectrum as well as the current--phase relationship in a single-channel SNS-junction, for any combination of effective Zeeman and spin--orbit fields and across the whole range of $T$, from a fully transparent to a fully opaque junction.
We found that our results exhibit the expected behavior as a function of all parameters, and that our expressions reproduce the known results in limiting cases.
The fact that our analytic expressions are most general presents the great advantage that they can be used to explore additional limits as well, by straightforward simplification.
Below, we will first consider the fully transparent limit $T=1$, which has been investigated before in the literature~\cite{Yokoyama2014}, and we will show how our expressions reproduce the expected results.
Then we will explore the experimentally relevant limit of $T \ll 1$, deriving relatively compact expressions for the spectrum and the supercurrent, which also allow us to find an analytic expression for the critical current through the junction.
We will derive compact analytic expressions for the current in the limiting cases of strong spin--orbit coupling and vanishing spin--orbit coupling.
Finally, we will show how setting $\alpha$ to a specific value highly reduces the complexity of the expressions, yielding compact results that are valid for any $0<T\leq 1$.
As an example, we give the results for $\alpha = 0$ (most scattering takes place at one of the interfaces) and $\alpha = 1/2$ (the scatterer is placed in the center of the wire).

\subsection{High-transparency limit, $T = 1$}

\begin{figure}[t]
    \centering
    \includegraphics[width=0.47\textwidth]{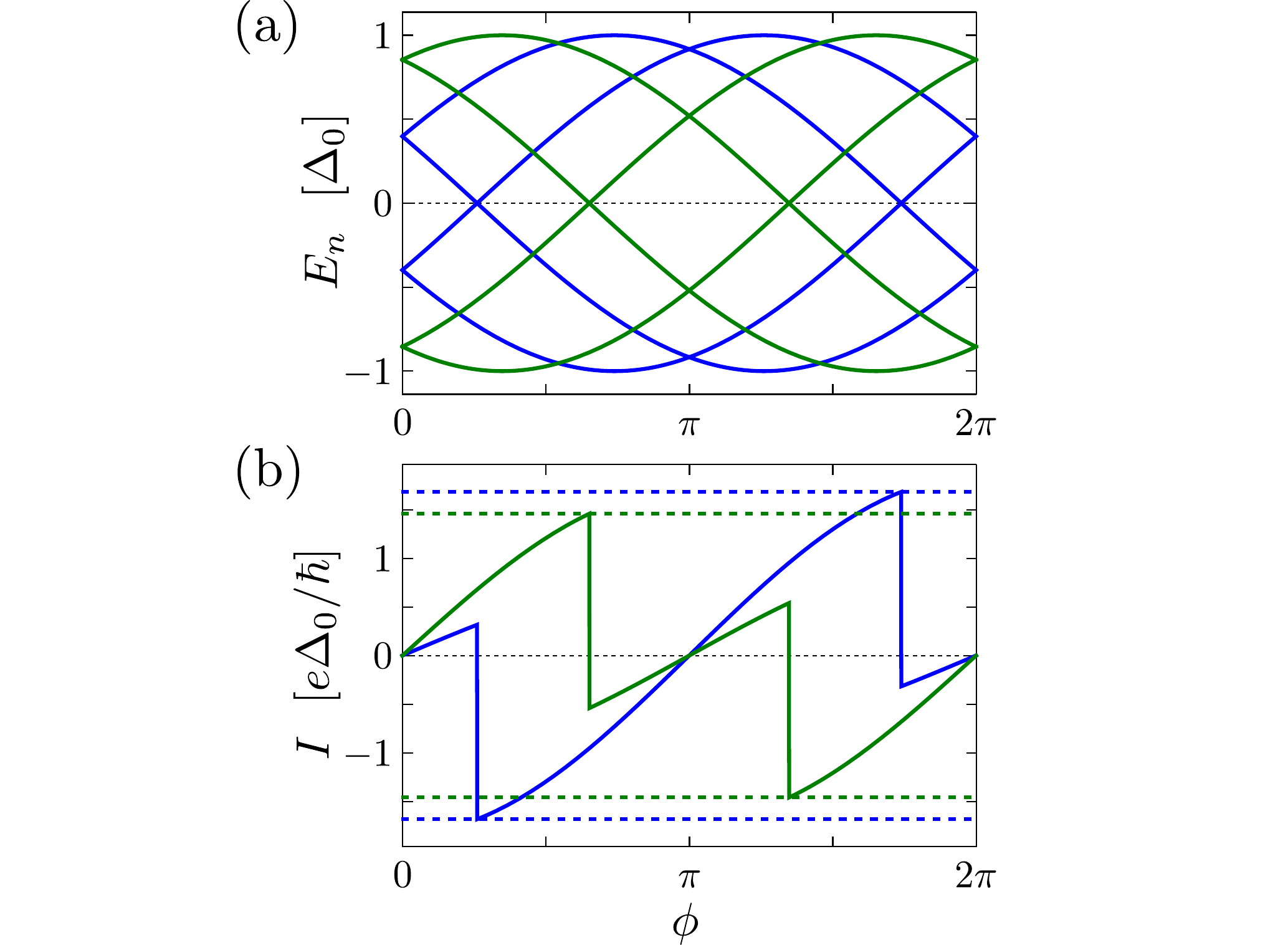}
    \caption{Calculated (a) bound-state energies and (b) supercurrent using Eqs.~(\ref{eq:energy_T1}) and (\ref{eq:current_T1}).
    Blue: $\chi_{\rm Z} = 0.4\,\pi$ and $\chi_{\rm so} = 0.2\,\pi$.
    Green: $\chi_{\rm Z} = 0.2\,\pi$ and $\chi_{\rm so} = 0.4\,\pi$.
    In both cases we set the angle between the fields to $\pi/4$.
    }
    \label{fig:T1}
\end{figure}

In the case of a fully transparent wire with $T=1$, the location of the scatterer becomes irrelevant and Eq.~(\ref{eq:energy}) can be simplified to
\begin{equation}\label{eq:energy_T1}
    E(\phi) = \pm \frac{\Delta_0}{\sqrt 2} \sqrt{1 + \cos(\phi\pm\zeta) },
\end{equation}
with the angle $0\leq \zeta\leq \pi$ defined through
\begin{equation}
     \cos(\zeta) = \cos(\chi_{+})\cos(\chi_{-}) - \cos(\theta)\sin(\chi_{+})\sin(\chi_{-}).\label{eq:expra}
\end{equation}
We decided to express the energies in the same form as in Eq.~(\ref{eq:energy}), but it is clear that this result is equivalent to $E(\phi) = \pm \cos [  \frac{1}{2}(\phi \pm  \zeta)]$, as, e.g., derived in Ref.~\cite{Yokoyama2014}.
In this fully-transparent limit the levels show the regular $\cos(\phi/2)$-behavior, with a phase shift equal to $\pm\zeta$, which is the result of the interplay of the effects of the Zeeman and spin--orbit fields on the phase of the propagating particles.
This is illustrated in Fig.~\ref{fig:T1}(a), where we plot the bound-state spectrum as given by Eq.~(\ref{eq:energy_T1}) with $\chi_{\rm Z} = 0.4\,\pi$ and $\chi_{\rm so} = 0.2\,\pi$ (blue curves) and $\chi_{\rm Z} = 0.2\,\pi$ and $\chi_{\rm so} = 0.4\,\pi$ (green curves), setting the angle between the two fields to $\pi/4$ in both cases.

The equilibrium supercurrent then follows as
\begin{equation}\label{eq:current_T1}
    I(\phi) =\frac{e\Delta_0}{2\hbar}  \sum_{\eta = \pm 1} \frac{\sin(\phi + \eta\,\zeta)}{|\cos[\frac{1}{2}(\phi+\eta\,\zeta)]|},
\end{equation}
cf.~the expressions given in Ref.~\cite{Yokoyama2014}.
Fig.~\ref{fig:T1}(b) shows the supercurrent following from this expression, for the same two sets of parameters as used in the left pane.
A straightforward inspection of this result yields that the corresponding critical current in the fully transparent case reads as
\begin{equation}\label{eq:icT1}
    I_c = \frac{e\Delta_0}{\hbar}[1+|\cos(\zeta)|],
\end{equation}
in line with the results presented in Ref.~\cite{Yokoyama2014}.
For comparison, we added the values of $\pm I_c$ predicted by Eq.~(\ref{eq:icT1}) to the bottom pane of Fig.~\ref{fig:T1} as horizontal dashed lines.

We thus see that our expressions reproduce the known results for the transparent case.
The unorthodox form of Eq.~(\ref{eq:energy_T1}), based on that of Eq.~(\ref{eq:energy}), allows for a more straightforward evaluation of the ground-state energy and thus the equilibrium supercurrent since the sign of each of the energies is obvious from the expression.

\subsection{Low-transparency limit, $T\ll 1$}

\begin{figure*}[t!]
    \centering
    \includegraphics[width=\textwidth]{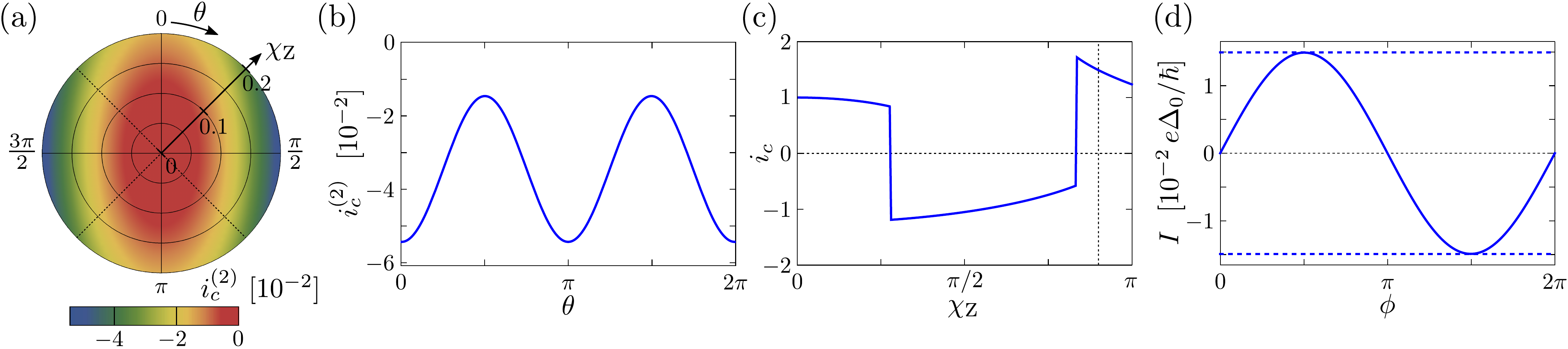}
    \caption{Critical current in the opaque regime.
    (a) Correction to the critical current in the opaque regime in the presence of strong spin--orbit coupling.
    We plot $i_c$ to leading order in $\chi_{\rm Z} / \chi_{\rm so}$ in the limit of $\chi_{\rm Z} / \chi_{\rm so}\ll 1$, as given by Eq.~(\ref{eq:ic2}), as a function of the angle $\theta$ between ${\bf B}_{\rm Z}$ and ${\bf B}_{\rm so}$ (polar angle) and $\chi_{\rm Z}$ (radial coordinate) for $\chi_{\rm so} = 0.4\,\pi$ and $\alpha = 0.4$.
    (b) Plot of the same correction $i^{(2)}_c$ for the same $\chi_{\rm so}$ and $\alpha$, at $\chi_{\rm Z} = 0.2$, i.e., at the outer circle of the plot shown in (a).
    (c) The critical current $i_c$ as a function of $\chi_{\rm Z}$ in the opposite limit of vanishing spin--orbit coupling, where we used $\alpha = 0.1$.
    (d) The current-phase relationship for $\chi_{\rm Z} = 0.9\,\pi$ [vertical dotted line in (c)], $\chi_{\rm so}=0$, $T=0.1$, and $\alpha = 0.1$. 
    }
    \label{fig:smallT}
\end{figure*}

For the case of a relatively opaque junction, we take $T\ll 1$ and thus expand Eqs.~(\ref{eq:energy}) and (\ref{eq:current}) to leading order in $T$.
This yields, as expected for low transparency, a sinusoidal current--phase relationship,
\begin{equation}
    I(\phi) = \frac{e\Delta_0}{\hbar} i_c T^2\sin(\phi),
\end{equation}
where the dependence of the current on the Zeeman and spin--orbit fields is captured by
\begin{equation}
    i_c = \sum_{\eta = \pm 1}
    \frac{q_2(1) + \eta \frac{2 q_4(\frac{1}{2}\sqrt 2) + 4 q_1(\frac{1}{2}\sqrt 2) q_2(\frac{1}{2}\sqrt 2)}{\sqrt{\frac{1}{2}+q_1(0)^2-q_3(0)}}}{\sqrt{2-2q_1(0)- 2\eta\sqrt{\frac{1}{2}+q_1(0)^2-q_3(0)}}},\label{eq:icsmallT}
\end{equation}
written again in terms of the coefficients $q_i(T)$ as given in App.~\ref{app:exp}.
Although thus cumbersome, this analytic expression for the critical current can be used to reveal useful insights in the dependence of the current on the Zeeman and spin--orbit fields and their relative orientation.

For example, for the case of a junction with strong spin--orbit coupling we can investigate perturbatively the effect of the application of a small Zeeman field in addition.
We thus expand Eq.~(\ref{eq:icsmallT}) to leading order in $\chi_{\rm Z}/\chi_{\rm so}$, assuming also $\chi_{\rm Z}\ll1$, yielding $i_c \approx 1 + i_c^{(2)}$ with
\begin{align}
    i_c^{(2)} = {} & {}
    - 2 \frac{\chi_{\rm Z}^2}{\chi_{\rm so}^2} \cos( \chi_{\rm so}) \sin( \alpha \chi_{\rm so}) \sin(\tilde \alpha\chi_{\rm so})\sin^2(\theta)
    \nonumber \\ {} & {}
    - 2 \chi_{\rm Z}^2 \tilde \alpha \alpha \cos^2(\theta),\label{eq:ic2}
\end{align}
where $\tilde \alpha = 1-\alpha$.
In Fig.~\ref{fig:smallT} we show the resulting second-order correction for $\chi_{\rm so} = 0.4\,\pi$ and $\alpha = 0.4$, (a) as a function of both $\theta$ and $\chi_{\rm Z}$ and (b) as a function of $\theta$ for $\chi_{\rm Z} = 0.2$.
We see that the (negative) correction is maximal for ${\bf B}_{\rm Z}$ and ${\bf B}_{\rm so}$ (anti)aligned and minimal when the two fields are perpendicular.
From Eq.~(\ref{eq:ic2}) it is indeed easy to see that the maximum correction $- 2 \chi_{\rm Z}^2 \tilde \alpha \alpha $ is reached at $\theta = 0,\pi$ and the minimum correction $-2(\chi_{\rm Z}/\chi_{\rm so})^2\cos( \chi_{\rm so}) \sin( \alpha \chi_{\rm so}) \sin(\tilde \alpha\chi_{\rm so})$ at $\theta = \frac{\pi}{2},\frac{3\pi}{2}$.
In an experiment, one could thus not only relate the directions of the applied ${\bf B}_{\rm Z}$ for which maxima and minima in the critical current are observed to the direction of the effective spin--orbit field, but one could also obtain information about the (probably phenomenological) parameter $\alpha$ and about $\chi_{\rm so}$, by following the development of the maxima and minima as a function of $\chi_{\rm Z}$.

We can also consider the opposite case of vanishing spin--orbit interaction, where the critical current $i_c$ becomes a function of $\alpha$ and $\chi_{\rm Z}$ alone,
\begin{align}
    i_c = \frac{\tan(2\alpha \chi_{\rm Z})|\cos (2\alpha \chi_{\rm Z})| - \tan(2\tilde \alpha \chi_{\rm Z})|\cos (2\tilde \alpha \chi_{\rm Z})|}{\sin( 2[\alpha - \tilde \alpha]\chi_{\rm Z})}.\label{eq:iczeroSOI}
\end{align}
In Fig.~\ref{fig:smallT}(c) we show the critical current following from Eq.~(\ref{eq:iczeroSOI}) for the case $\alpha = 0.1$.
As an example, we plot in Fig.~\ref{fig:smallT}(d) the current--phase relationship for $\chi_{\rm Z} = 0.9\,\pi$, $\chi_{\rm so}=0$, $T=0.1$, and $\alpha = 0.1$, as calculated from Eq.~(\ref{eq:current}), where the blue dashed lines represent the critical current as given by Eq.~(\ref{eq:iczeroSOI}).
The vertical dotted line in Fig.~\ref{fig:smallT}(c) indicates the $\chi_{\rm Z}$ that was used in Fig.~\ref{fig:smallT}(d).
We note that the sign changes in $i_c$ correspond to $0$--$\pi$ transitions in the junction.

\subsection{Fixed position of the scatterer}

Finally, we show how the general expression given for the bound-state spectrum (\ref{eq:energy}) simplifies when the location of the scatterer is fixed.
Here we choose the particularly simple values of $\alpha = 0$ and $\alpha = 1/2$, but similar simplifications can be made for any value of $\alpha$.

When $\alpha = 0$, i.e., the scattering takes places mostly at one end of the wire, then the coefficients $q_i$ as given in App.~\ref{app:exp} simplify considerably, yielding
\begin{equation}\label{eq:energyAlpha0}
        E(\phi) = \pm \frac{\Delta_0}{\sqrt 2} \sqrt{v_0\pm \sqrt{w_0}},
\end{equation}
with now
\begin{align}
    v_0 = {} & {} 1 + \cos(\zeta)[\cos(\zeta) R^2+T^2\cos(\phi)],\\
    w_0 = {} & {} \sin(\zeta)^2\big( 1- [\cos(\zeta) R^2+T^2\cos(\phi)]^2 \big),
\end{align}
using the same angle $\zeta$ as before [see Eq.~(\ref{eq:expra})] and the notation $R^2 = 1-T^2$.
Defining
\begin{equation}
    \cos(\kappa) = \cos(\zeta) R^2+T^2\cos(\phi),
\end{equation}
with $0\leq\kappa\leq \pi$, we see that the energies become
\begin{equation}
        E(\phi) = \pm \Delta_0 \cos[ \tfrac{1}{2}(\kappa\pm\zeta)],
\end{equation}
where the phase-dependence is implicit via the angle $\kappa$.
This allows us to write a compact expression for the equilibrium supercurrent as well,
\begin{equation}
    I(\phi) = \frac{e\Delta_0}{2\hbar}  \sum_{\eta = \pm 1} \frac{T^2 \sin(\kappa + \eta\,\zeta)\sin(\phi)}{|\cos[\frac{1}{2}(\kappa+\eta\,\zeta)]\sin(\kappa)|},\label{eq:currentalpha0}
\end{equation}

Another simple choice for $\alpha$ that leads to relatively compact expressions is $\alpha = 1/2$, i.e., the scatterer placed in the center of the wire.
In this case, the energy reads again as
\begin{equation}\label{eq:energyAlpha0.5}
        E(\phi) = \pm \frac{\Delta_0}{\sqrt 2} \sqrt{v_\frac{1}{2}\pm \sqrt{w_\frac{1}{2}}},
\end{equation}
with now
\begin{align}
        v_\frac{1}{2} = {} & {} 1  + \cos(\zeta)[R^2 + T^2\cos (\phi) ] - 2\sin^2(\lambda) R^2,\\
        w_\frac{1}{2} = {} & {} 16 \sin^2(\zeta/2) T^2\sin^2(\phi/2)
        \\ {} & {} \hspace{1em} \times 
        \left\{ \cos^2(\zeta/2) [ 1-T^2\sin^2(\phi/2)] - \sin^2(\lambda) R^2 \right\},\nonumber
\end{align}
using the same definitions as before and
\begin{align}
     \sin(\lambda) = \sin (\chi_{+}/2) \sin(\chi_{-}/2)\sin(\theta).
\end{align}
In this case, the current--phase relationship can still be straightforwardly calculated, but does not reduce to an expression as elegant as Eq.~(\ref{eq:currentalpha0}).

Other values of $\alpha$ can also straightforwardly be explored using the explicit results given in App.~\ref{app:exp}; the resulting expressions will most likely be of similar complexity as the ones for $\alpha = 0, 1/2$.

\section{Conclusion}\label{sec:conclusion}

To conclude, we presented analytic expressions for the Andreev bound-state energies and supercurrent in a single-channel semiconductor-based SNS-junction.
We took into account the effects of finite spin--orbit interaction and a Zeeman splitting, allowing the two associated fields to have any desired orientation, and we included a single scatterer that can be arbitrarily located in the normal region.
We first corroborated our results by comparing them to all limiting cases for which analytic expressions are known.
Then we demonstrated how our general expressions can be simplified in several additional limits, such as that of small and large spin--orbit coupling in the low-transparency case, resulting in insightful expressions that could be helpful for disentangling spin--orbit and Zeeman effects in experiment.

We gratefully acknowledge financial support via NTNU's Onsager Fellow program.

\appendix

\begin{widetext}
\section{Explicit expressions for $q_i$}
\label{app:exp}

The four coefficients $q_i$ read explicitly as
\begin{align}
q_1(T) = {} & {} R^2 \sin(\theta)^2 [ \sin(\alpha \chi_+)^2 \sin(\alpha \chi_-)^2 + \sin(\tilde \alpha \chi_+)^2 \sin(\tilde \alpha \chi_-)^2 ] - \tfrac{1}{2} R^2 [f_2 (2\alpha\chi_+,2\alpha \chi_-) + f_2(2\tilde \alpha\chi_+,2\tilde \alpha\chi_-)],\\
q_2(T) = {} & {} T^2 f_2(\chi_+,\chi_-),\\
q_3(T) = {} & {} \tfrac{1}{4}R^2 T^2 \sin(\theta)^2 [f_2 (2 \alpha  \chi_+,-2 \tilde\alpha \chi_-)+f_2(2 \alpha  \chi_-,-2 \tilde\alpha \chi_+)]
-\tfrac{1}{8} R^4 \sin(\theta)^4-4 T^4 \sin(\theta)^2 \sin(\chi_+)^2 \sin(\chi_-)^2
\nonumber\\ {} & {} 
+\tfrac{1}{4}R^4 \sin(\theta)^2 [2+f_2 (2 \alpha  \chi_+,-2 \beta \chi_-)+f_2(2 \tilde\alpha \chi_+,2 \beta \chi_-)+f_2(2 \alpha  \chi_-,-2 \beta \chi_+)+f_2(2 \tilde\alpha \chi_-,2 \beta \chi_+)
\nonumber\\ {} & {} 
\hspace{6em}-f_2(2 \alpha  \chi_+,2 \alpha  \chi_-)-f_2(2 \tilde\alpha \chi_+,2 \tilde\alpha \chi_-)]
\nonumber\\ {} & {} 
+\tfrac{1}{4}R^2 \sin(\theta)^2 [ f_2(2 \chi_+,2 \alpha  \chi_-)+f_2(2 \chi_+,2 \tilde\alpha \chi_-)+f_2(2 \chi_-,2 \alpha  \chi_+)+f_2(2 \chi_-,2 \tilde\alpha \chi_+)
\nonumber\\ {} & {}
\hspace{6em} -f_2(2 \alpha  \chi_+,2 \alpha  \chi_-)-f_2(2 \tilde\alpha \chi_+,2 \tilde\alpha \chi_-)] 
\nonumber\\ {} & {}
+R^2 T^2 f_6 (2 \chi_+,2 \chi_-)+\tfrac{1}{2} R^4 [ f_8(2 \beta \chi_+,2 \beta \chi_-)+f_8(2 \chi_+,2 \chi_-)+f_8(0,0)] +\tfrac{1}{2} T^4 f_2(2 \chi_+,2 \chi_-)
\nonumber\\ {} & {} 
+\tfrac{1}{4}\left(T^2+R^2 \sin(\theta)^2 [2 \sin(\alpha  \chi_+)^2+2 \sin(\tilde\alpha \chi_+)^2-\sin(\chi_+)^2-\sin(\beta \chi_+)^2]\right)
\nonumber\\ {} & {} 
\hspace{3em}\times \left(T^2 + R^2 \sin(\theta)^2 [2 \sin(\alpha  \chi_-)^2+2 \sin(\tilde\alpha \chi_-)^2-\sin(\chi_-)^2-\sin(\beta \chi_-)^2]\right)
\nonumber\\ {} & {}
-\tfrac{1}{4}\left[T^2+\tfrac{1}{2} R^2 \sin(\theta)^2 \cos (2 \beta \chi_+)\right] \left[T^2+\tfrac{1}{2} R^2 \sin(\theta)^2 \cos (2 \beta \chi_-)\right]
\nonumber\\ {} & {}
-\tfrac{1}{4}\left[T^2-\tfrac{1}{2} R^2 \sin(\theta)^2 \cos (2 \chi_+)\right] \left[T^2-\tfrac{1}{2} R^2 \sin(\theta)^2 \cos (2 \chi_-)\right]+\tfrac{1}{4}\left[T^2-\tfrac{1}{2} R^2 \sin(\theta)^2\right]^2,\\
q_4(T) = {} & {} T^2R^2 \left( 2f_2(\beta \chi_-,\beta \chi_+) - \sin(\theta)^2 [ \cos(\chi_+) - \cos(\beta\chi_+)][\cos(\chi_-) - \cos(\beta\chi_-)] \right),
\end{align}
where we used the function
\begin{equation}
f_n (x,y) = \cos \left(\tfrac{1}{2}\theta\right)^n \cos(x+y) + \sin \left(\tfrac{1}{2}\theta\right)^n \cos(x-y),
\end{equation}
and the notation $\tilde \alpha = 1-\alpha$, $\beta = \tilde \alpha - \alpha$, and $R = \sqrt{1-T^2}$.

\end{widetext}


%

\end{document}